%% file: main.tex
\documentclass[%
 reprint,
superscriptaddress,
 aps,
justification=justified
]{revtex4-2}

\usepackage{graphicx}
\usepackage{bm}
\usepackage{hyperref}
\usepackage[mathlines]{lineno}
\usepackage{url}
\usepackage{color}
\usepackage{graphicx}
\usepackage{mathtools}
\usepackage{float}
\usepackage[export]{adjustbox}[2011/08/13]
\usepackage{gensymb} 
\usepackage{indentfirst}
\usepackage{mathrsfs}
\usepackage{cuted}
\usepackage{wrapfig}

\usepackage{multirow}
\usepackage[utf8]{inputenc}

\renewcommand{\figurename}{Fig.}

\begin{document}

\preprint{APS/123-QED}

\title{A low-threshold ultrahigh-energy neutrino search with the Askaryan Radio Array}

\input{ara_revtex_institutes.tex}
\input{ara_revtex_authors.tex}

\date{\today}

\begin{abstract}

In the pursuit of the measurement of the still-elusive ultrahigh-energy (UHE) neutrino flux at energies of order EeV, detectors using the in-ice Askaryan radio technique have increasingly targeted lower trigger thresholds. This has led to improved trigger-level sensitivity to UHE neutrinos. 
Working with data collected by the Askaryan Radio Array (ARA), we search for neutrino candidates at the lowest threshold achieved to date, leading to improved analysis-level sensitivities. A neutrino search on a data set with 208.7~days of livetime from the reduced-threshold fifth ARA station is performed, achieving a 68\% analysis efficiency over all energies on a simulated mixed-composition neutrino flux with an expected background of $0.10_{-0.04}^{+0.06}$ events passing the analysis. We observe one event passing our analysis and proceed to set a neutrino flux limit using a Feldman-Cousins construction. We show that the improved trigger-level sensitivity can be carried through an analysis, motivating the Phased Array triggering technique for use in future radio-detection experiments. We also include a projection using all available data from this detector. Finally, we find that future analyses will benefit from studies of events near the surface to fully understand the background expected for a large-scale detector.

\end{abstract}

\maketitle

\clearpage
\newpage
\mbox{~}
\clearpage
\newpage

\section{Introduction}

Neutrinos are a powerful tool for understanding the Universe at the highest energies. While other messenger particles such as cosmic rays and gamma rays are either deflected or absorbed on the way to Earth after being created by their astrophysical sources, neutrinos point back to their sources and rarely interact, making them unique messengers for understanding distant or dense astrophysical sources. Additionally, we expect cosmogenic neutrinos, created via interactions between cosmic rays and the cosmic microwave background \cite{G_ZK}; measuring the cosmogenic neutrino flux will help answer fundamental questions about cosmic ray composition at the highest energies. The IceCube experiment has measured an astrophysical neutrino flux up to 10~PeV \cite{IceCube_limit1}\cite{IceCube_limit2} and has early hints of a potential first source \cite{IceCube:2018dnn}\cite{IceCube:2018cha}. However, because the neutrino flux falls as energy increases, experiments searching for neutrinos above 10~PeV have needed to build larger detectors that observe many cubic kilometers of ice to have a chance of detecting these rare events.
\begin{figure}[ht]
\centering
\includegraphics[width=0.49\textwidth]{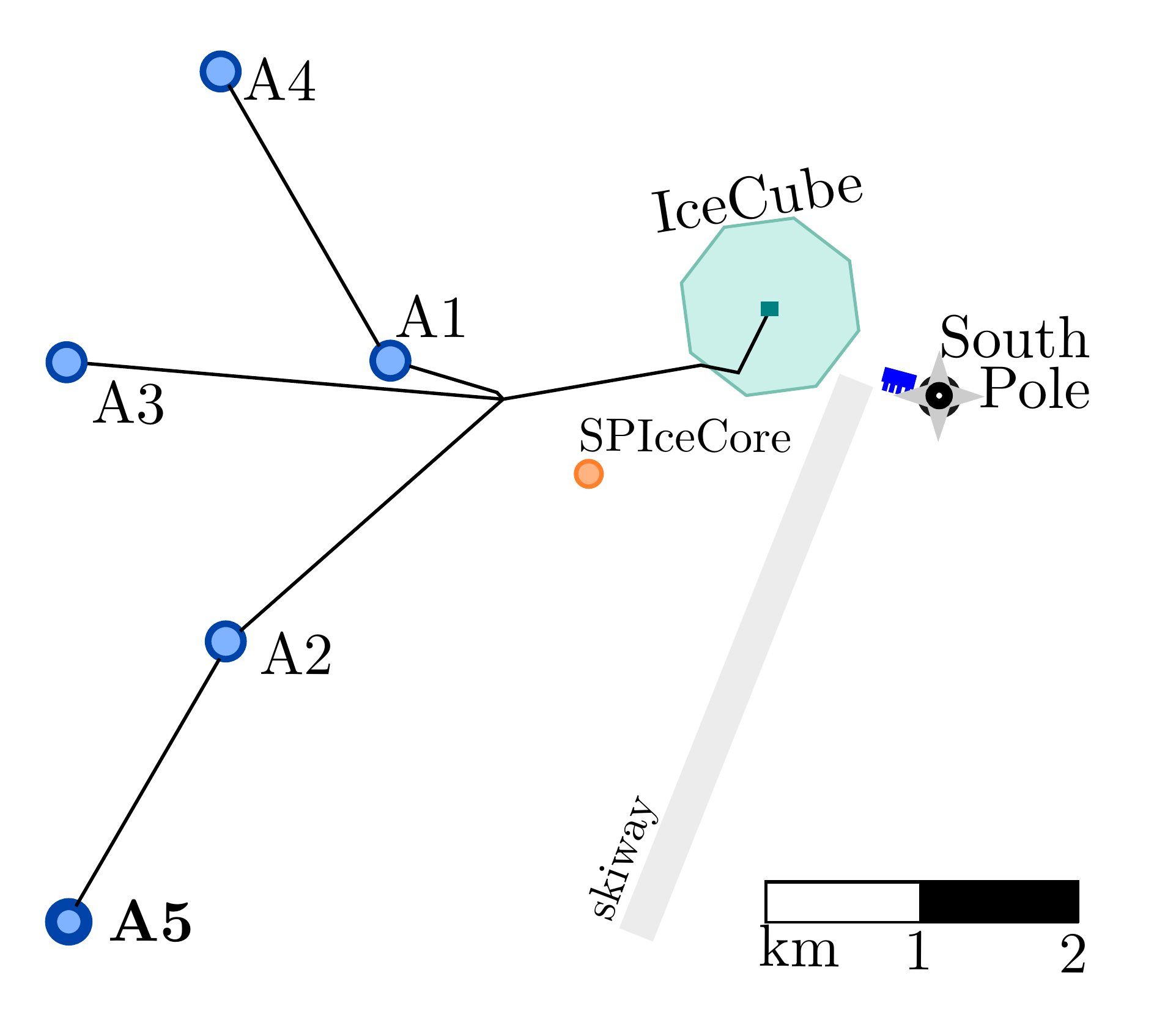}
\includegraphics[width=0.49\textwidth]{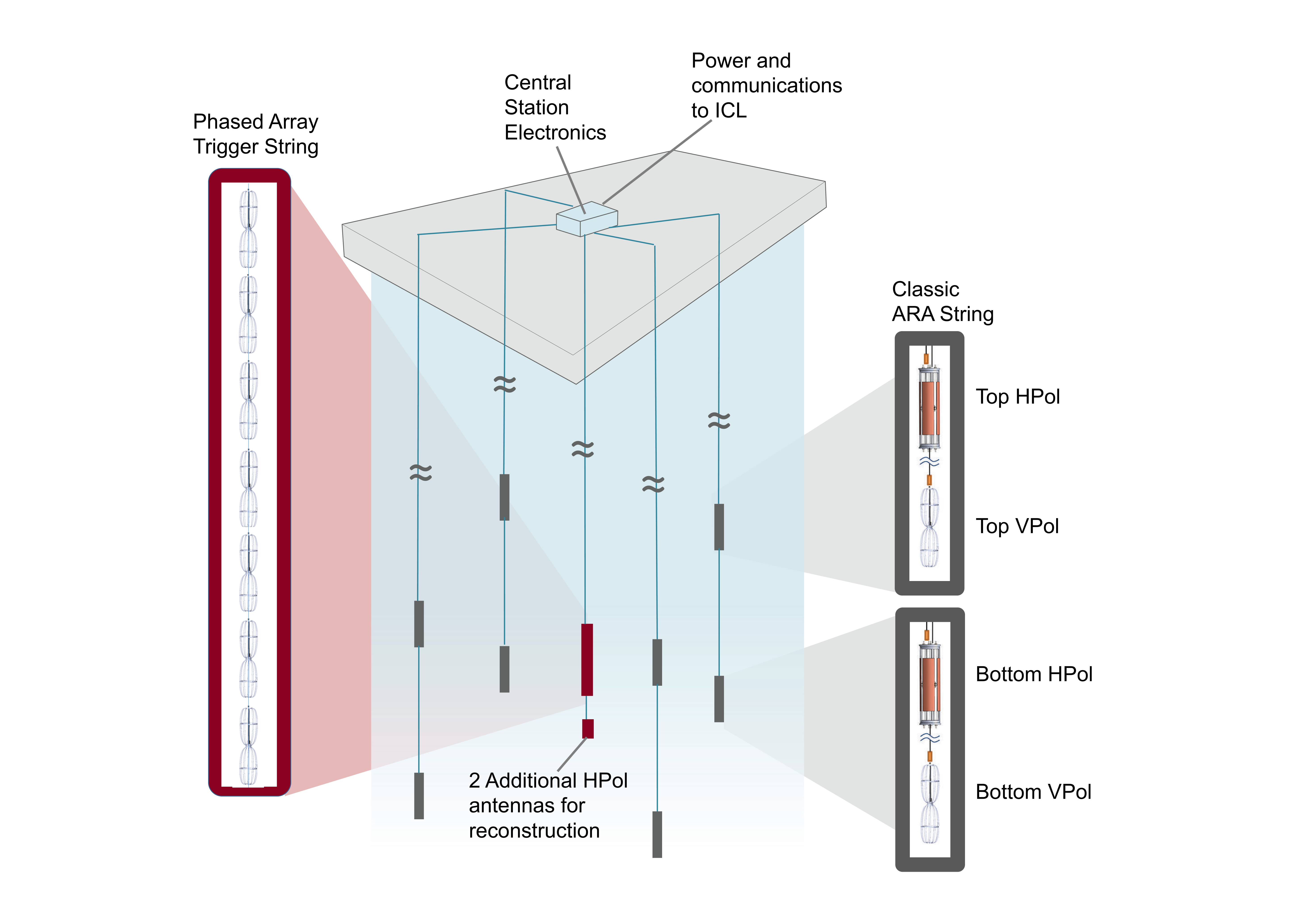}
\caption{Top: Bird's eye view of the ARA instrument relative to South Pole Station. Black lines indicate cable connections for power and network. This work focuses on ARA Station 5 (A5), the only station with an additional phased array trigger. Bottom: a schematic of the A5 detector. Gray antennas are the baseline ARA antennas while the red antennas are the Phased Array antennas.}
\label{fig:ara5_layout}
\end{figure}
One such experiment, the Askaryan Radio Array (ARA), takes advantage of the Askaryan effect, first proposed in 1962 by G. Askaryan \cite{Askaryan:1962hbi} and which influences the design of many neutrino experiments at high energies \cite{ARIANNA} \cite{ANITA_IV} \cite{RNOG_2021} . When a neutrino interacts in matter, a particle shower occurs, which develops a negative charge excess of $\mathcal{O}(\mathrm{20\%})$ due to scattering with the electrons in the medium, and positrons in the shower annihilating with electrons in the matter. 
This negative charge excess causes two types of radiation to occur: bremsstrahlung-like radiation, caused by the transient nature of the net charge, and Cerenkov-like radiation, caused by the net charge moving faster than the speed of light in the interaction media. For wavelengths greater than the apparent lateral width of the shower (determined by the Moli\`ere radius of the material and the viewing angle), the radiation is coherent. 

Ice is one medium in which Askaryan radiation can be observed, due to its transparency to radio-frequency radiation\cite{AskaryanIce}; it has also been observed in the lab in sand, rock salt, polyethylene, and the atmosphere \cite{AskaryanSand} \cite{AskaryanRockSalt} \cite{AskaryanPolyethylene} \cite{AskaryanAtmosphere}. In ice, the Moli\`ere radius is approximately 10~cm,  which leads to coherent radiation and consequently an increase in power at radio wavelengths. As an additional benefit, Antarctic ice has a radio attenuation length of $\mathcal{O}(\mathrm{1~ km})$ \cite{ barwick_besson_gorham_saltzberg_2005}, which allows detectors to survey many cubic kilometers of ice with relatively sparse instrumentation.

In this paper, we report on the ARA Station 5 (A5) instrument, including the new Phased Array detector (also called NuPhase \cite{PhasedArrayInstrument}), and we present the first neutrino search of data triggered by the Phased Array. Demonstrating that low-threshold events can pass selection in a neutrino search is the crucial next step in demonstrating the improvement in effective volumes brought about by the phased array, and in motivating the design of future Phased Array triggers on other experiments \cite{IceCubeGen2_White} \cite{RNOG_2021}.

\section{Experimental Apparatus and Data Set} 
\subsection{The ARA Detector} \label{detector_section}
The ARA instrument (Figure~\ref{fig:ara5_layout}) consists of five independent stations, arranged in a triangular grid with neighboring stations separated by two kilometers and all deployed near South Pole Station in Antarctica. Each station consists of sixteen antennas installed in holes drilled in the ice to a maximum depth of 200~m. Both vertically-polarized (VPol) and horizontally-polarized (HPol) antennas are used so that polarization of the incoming signal can be measured. ARA is sensitive to electromagnetic radiation of the frequency range 150-850~MHz \cite{ARAtestbed}.

The newest station, A5, was built in 2018 and has two separate but connected detectors, as seen in Figure~\ref{fig:ara5_layout}: one, the typical ARA station design described above, hereafter referred to as the baseline instrument; and two, an additional central antenna string with its own triggering instrument called the Phased Array \cite{PhasedArrayInstrument}. Unlike the baseline design, which has antennas deployed in pairs of VPol and HPol antennas separated by a vertical distance between 20-30 m, the Phased Array instrument consists of seven VPol antennas and two HPol antennas, deployed with compact 1-2 m separation \cite{PhasedArrayConcept}. While the Phased Array and the baseline system are connected, they are mostly separate instruments, each with their own Data Acquisition (DAQ) system.

The baseline A5 trigger requires three antennas of the same polarization to record their integrated power over a 25~ns time window as greater than five times the ambient thermal noise level, within a coincidence of approximately 170~ns. The baseline A5 instrument records this type of trigger at a rate of $\sim$~6~Hz, with an additional software trigger at a rate of 1~Hz. More details on the ARA electronics can be found in \cite{ARA23}.

The Phased Array instrument, described in deatil in \cite{PhasedArrayInstrument}, is designed with a beamforming trigger, a technique used often by radio telescope experiments \cite{LOFAR}. Incoming radio waves arrive at the antennas of the one-dimensional Phased Array in a specific order depending on the incoming zenith angle of the signal; thus, for distant, plane-wave-like signals, the time difference between the pulses on each channel are directly related to the arrival angle of the incoming signal. By defining 15 pre-determined directions (often called ``beams") based on the expected time delays seen at the antennas, signals are delayed appropriately and then summed together prior to the trigger. The power in each beam is monitored in 10~ns interleaved time intervals, with the trigger threshold continuously adjusted to meet a 0.75~Hz rate for each of the 15 beams for a global trigger rate of 11~Hz. Impulsive signals add coherently for the beam that they are associated with, while noise on average do not, lowering the trigger threshold compared to a baseline ARA station. The 50\% trigger efficiency point for the Phased Array instrument occurs at a single-antenna voltage-based signal-to-noise ratio (SNR) of 2.0, compared to an SNR of 3.7 for the baseline ARA trigger; the full definition of SNR can be found in \cite{PhasedArrayInstrument}. 

If the Phased Array triggers on an event, it externally triggers the baseline ARA system at A5. The baseline system may or may not have independently triggered on the event due to the baseline system having a higher trigger threshold. In normal operations, up to 98\% of Phased Array triggered events have a matching baseline event. If the baseline system triggers on an event, it does not externally trigger the Phased Array instrument.

This analysis only considers events triggered by the Phased Array instrument. As of the 2018-2019 deployment season, the output of one of the baseline ARA channels was split and connected to both the baseline ARA DAQ and the Phased Array DAQ. Because of this common channel, the timing of events that occur in both instruments can be synchronized and all available channels at A5 can be used to reconstruct the events.

\subsection{Livetime}
A5 was installed in the 2017-2018 deployment season and has been operational since that time. For this report, we only analyze data taken throughout the 2019 calendar year, to take advantage of the common channel described above. 

Within this dataset, approximately 3.5 months are not usable due to the Phased Array instrument overheating when operated in the warmest months. This overheating can be solved in future versions of a Phased Array system by adding additional heat-sinking measures or designing a system that consumes less power \cite{RNOG_2021}. Additionally, starting in October 2019 the baseline ARA system suffered a USB failure on its triggering and readout interface board, causing a loss in livetime of approximately 2 months. This leaves 208.7 days of total livetime in which both the Phased Array and the baseline system were taking data at the same time during 2019. The deadtime of the Phased Array system is estimated to be 0.12\% and has a negligible impact on the livetime.

\begin{figure}[ht]
\centering
\includegraphics[width=\columnwidth]{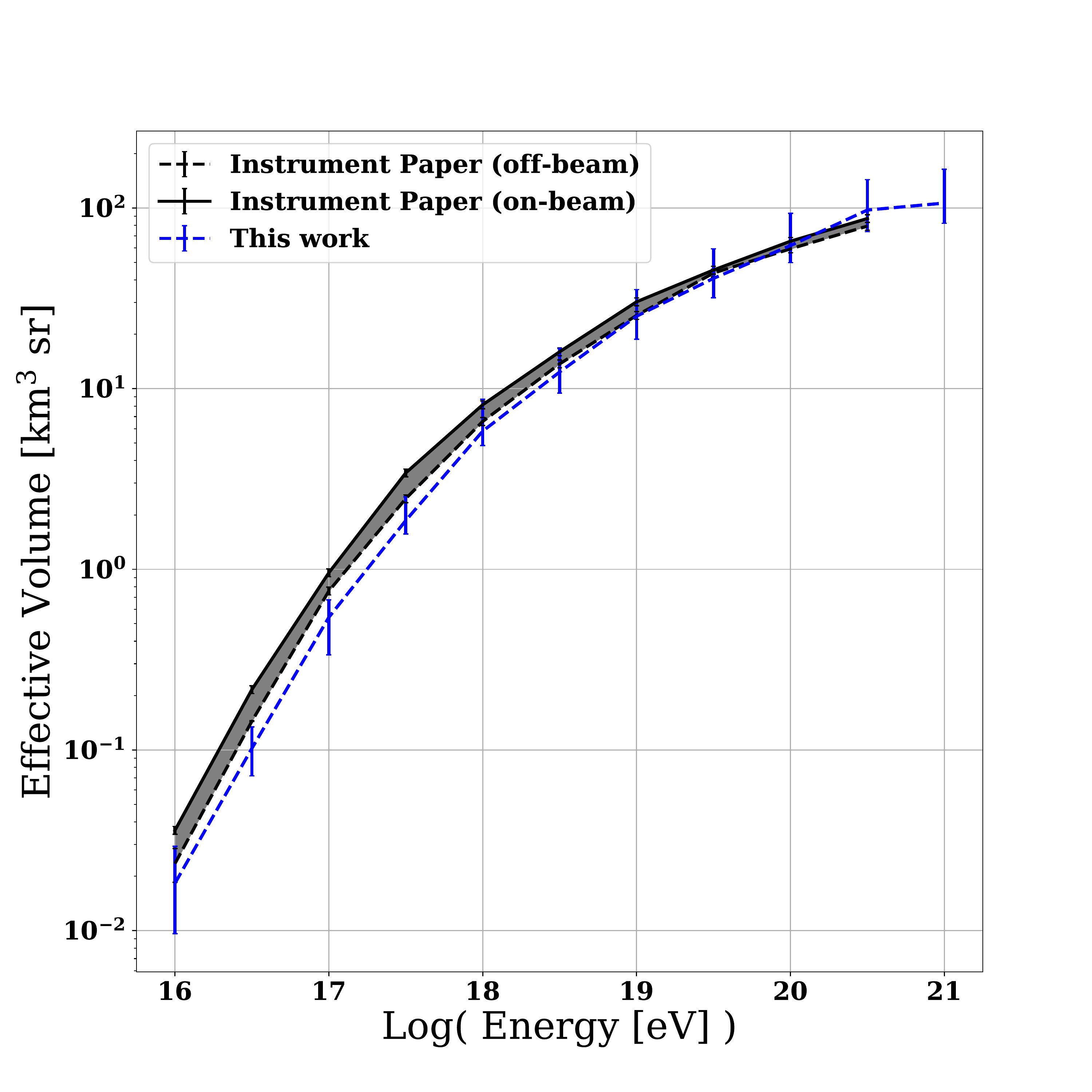}
\caption{A comparison of the effective volume of the simulated Phased Array trigger, compared to the previously published effective volume \cite{PhasedArrayInstrument}. The uncertainties for the blue curve include systematic uncertainties, which are described in Section \ref{uncertainties}.}
\label{fig:EffectiveVolume}
\end{figure}

We blind the data using the data prescaling method discussed by Klein and Roodman \cite{KleinRoodman}, unblinding 10\% of the full data set to select passing event criteria and extrapolate background estimates. The events comprising this 10\% data set are selected from all events by ordering events by their trigger time, breaking the data into groups of 10~events, and selecting one event at random from each group. We also unblind all data taken during one full 24-hour period to monitor any short-term cyclical behavior. The 10\% sample is not included in the final result.

While the analysis focuses on events triggered by the Phased Array, the baseline A5 instrument is used in three ways. First, the baseline instrument was used to calibrate the location of the Phased Array relative to the other landmarks at the South Pole \cite{ICRC_araCalib}. Second, the common channel that was connected to both the Phased Array and the baseline instrument was used in the analysis. Third, the baseline instrument was used to point surface events both in the 10\% sample and in the unblinded sample.

\begin{figure*}[htp]
\centering
\includegraphics[width=1.0\textwidth,center]{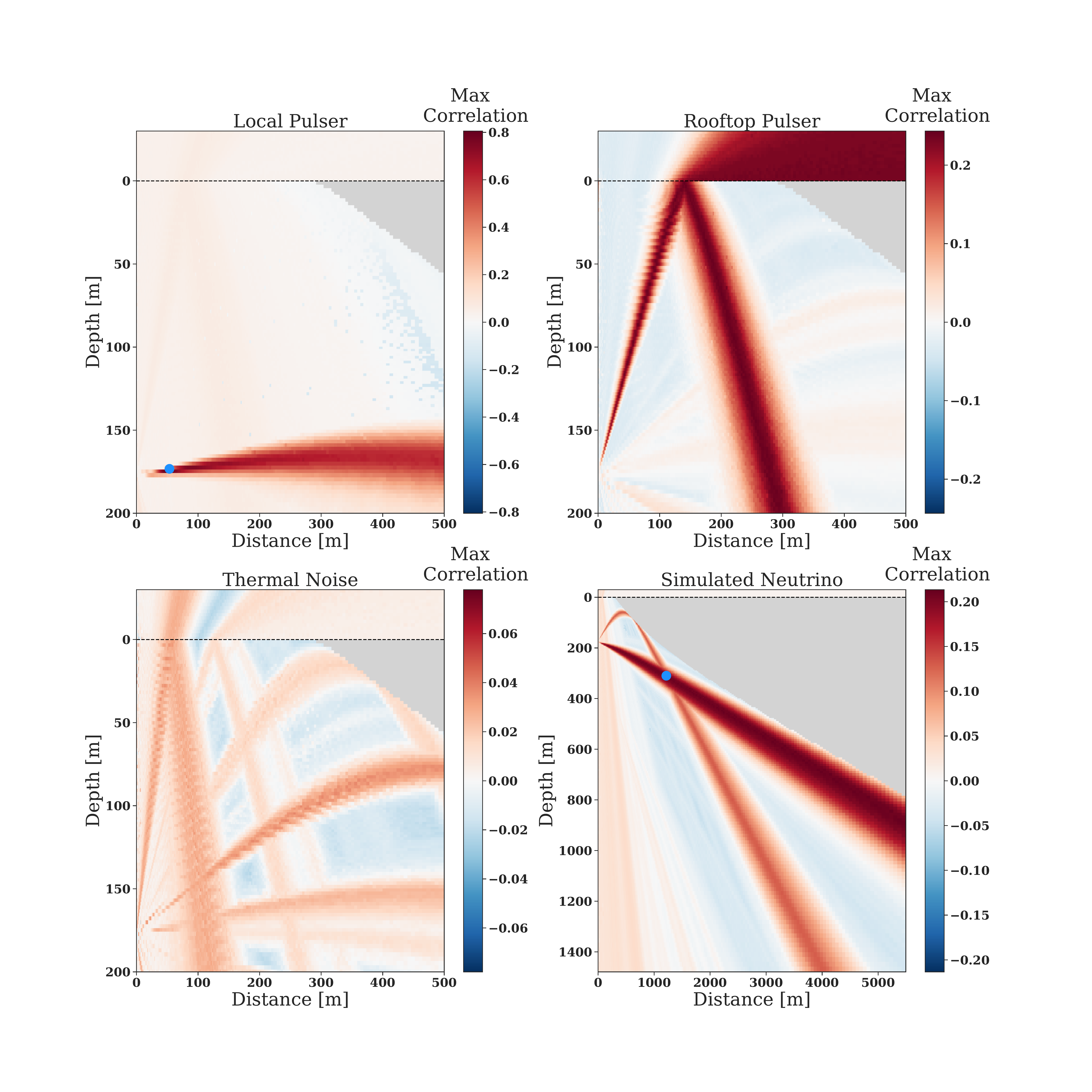}
\caption{Four example correlation maps created using only the antennas from the Phased Array instrument, centered at a depth of 170 m. The blue dots indicate the true location of the source, where applicable; the rooftop pulser location is out of frame, about 4 km away. The gray area is disallowed based on ray optics using the exponential firn model. The color map is scaled differently for each map to match the maximum correlation value for that event.} 
\label{fig:correlationMap}
\end{figure*}

\subsection{Simulation}

A modified version of the AraSim simulation package is used to simulate the entire A5 instrument, including both the baseline ARA channels and the Phased Array trigger string \cite{AraSim}. The trigger is implemented in the simulation following the methodology laid out in \cite{PhasedArrayInstrument}. First, the SNR and the incoming angle at the antenna array are calculated for each event. Second, each event is assigned a probability of passing the trigger based on two in-situ measurements: the trigger efficiency as a function of SNR, and the efficiency as a function of how ``on-beam" or ``off-beam" the incoming direction is with respect to the 15 beam directions of the trigger system discussed in Section \ref{detector_section}. For each event, a random number is generated between 0 and 1, and the event passes if its assigned probability is greater than the random number. The resulting simulated trigger matches the in-situ measurements of trigger efficiency observed in the Phased Array instrument. A comparison is shown in Figure \ref{fig:EffectiveVolume} between the effective volume for this analysis, calculated using the same Monte Carlo counting methods described in \cite{ARA23}, and the effective volume calculated in the Phased Array instrument paper \cite{PhasedArrayInstrument}.

Event selection in data analysis is optimized using simulated data sets, in which realistic simulation of the neutrino flux traversing the detector is crucial. We chose to simulate neutrinos using a flux based on a mixed composition of galactic cosmic rays and an optimistic model of source evolution \cite{Kotera:2010yn}. This simulation set includes neutrinos at all energies and is used to calculate analysis efficiency at each analysis stage. Separate data sets are simulated for different neutrino energies in half-decade energy bins from $10^{16}$ eV to $10^{21}$ eV to determine the analysis efficiency as a function of energy. 

\section{Ice Modeling and $R$-$z$ Plane Mapping} \label{IceModel}

The index of refraction of the ice changes as a function of depth due to the compacting of ice layers over time \cite{kravchenko_besson_meyers_2004}, leading to radio waves traveling in non-straight paths through the ice. This effect is largest in the upper few hundred meters of ice, the region called the firn. We estimate a model for the changing index of refraction by combining calibration data taken from both the local pulser, approximately 50 m away from the Phased Array and -170 m deep, and a pulser dropped in the SPIceCore borehole, approximately four kilometers away and between 900-1450 m in depth. The best fitting index of refraction model for the firn is found to be

\begin{equation}\label{eq:1}
n(z) = 1.780 - 0.454 e^{(-0.0202)\times \frac{z}{\text{1 m}} } ,
\end{equation}

\noindent where $z$ is the depth in meters. This is determined by starting with an existing internal ice model and modifying only the parameter in the exponent, using least sqaures regression to compare the predicted time delays from the model to the measured time delays between channels of the Phased Array. The ice model is specifically tuned to match only direct pulses from the SPIceCore calibration pulser, and did not fit for refracted pulses. This means the ice model likely describes deep ice most accurately. 
Using this model, the possible paths for a radio signal can be calculated for any source following Fermat's principle. For in-ice sources, this includes direct paths between the interaction and the detector, and refracted paths, in which the signal path curves above the detector before curving back down, made possible by the changing index of refraction.  By comparing the different path lengths for each antenna, expected time delays $\tau(R, z)$ between pulses on different channels can also be calculated for any given source as a function of the radial distance $R$ and the depth $z$ of that source. These expected time delays are characteristics of the environment, defined by the locations of the antennas relative to each other and the ice model.

For a given event, each antenna records a measure of voltage as a function of time, called a waveform. The cross correlation between the waveform of the $i$th antenna and the waveform of the $j$th antenna is calculated as a function of time delay $\tau$ using the correlation function

\begin{equation}
 \mathrm{Corr}_{i,j}(\tau)  = \sum_{t=0}^{N-1} \frac{V_i(t) V_j(t+\tau)}{\sigma_i \sigma_j}, 
\end{equation}

\noindent where $V_i(t)$ and $V_j(t+\tau)$ are the waveforms and $\sigma_i$ and $\sigma_j$ are the variances of the waveforms of the $i$th and $j$th channels, respectively. The location of the maximum of $\mathrm{Corr}_{i,j}(\tau)$ corresponds to the time delay that best aligns the signals. A map can be created describing the average correlation at every ($R$, $z$), described by

\begin{equation}
 C(R, z)  = \frac{\sum_{i=1}^{n_{ant}-1} \sum_{j=i+1}^{n_{ant}}\mathrm{Corr}_{i,j}[\tau(R, z)]}{\binom{n_{ant}}{2}}, 
\end{equation}

\noindent where, for the Phased Array, $n_{ant}=7$ is the number of antennas and $\binom{n_{ant}}{2}$ is the binomial coefficient. For each event, three correlation maps are created, corresponding to the $\tau(R, z)$ that describe the direct path, $\tau(R, z)$ for the refracted path, and the average between the two. For the plots in Figure \ref{fig:correlationMap}, the maximum between the direct and refracted maps is plotted. 
\begin{figure*}[htp]
  \centering

  \includegraphics[width=1.3\textwidth,center]{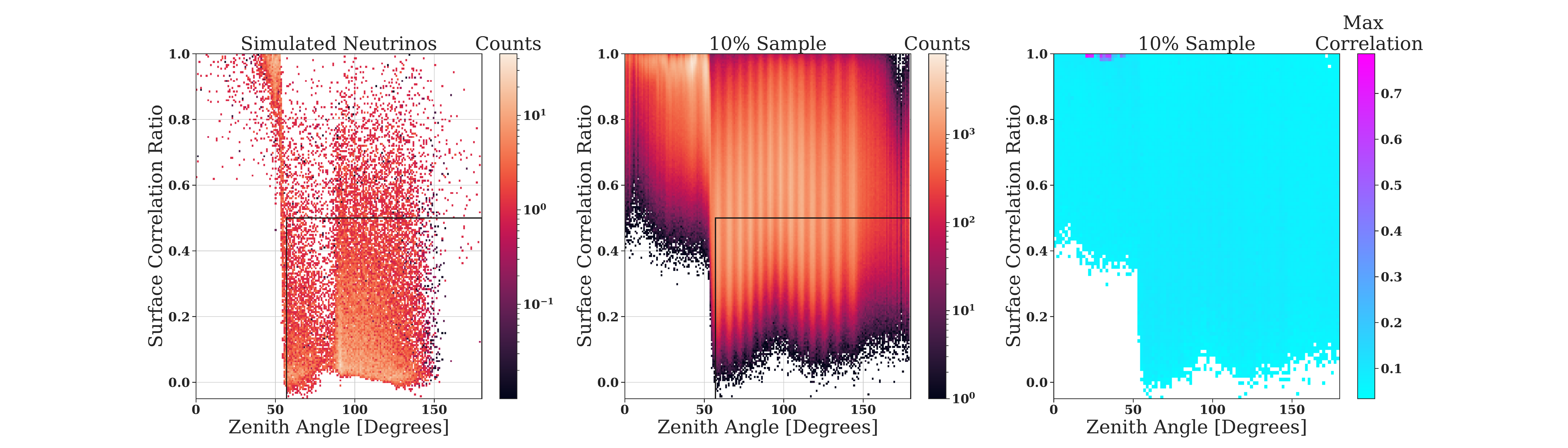}
  \caption{The distribution of the reconstructed zenith angle at the Phased Array vs.\@ the Surface Correlation Ratio for the simulated neutrino sample (left) and the 10\% sample (middle). The black box designates the signal region. The population in the upper left hand corner of simulated neutrinos corresponds to deep neutrinos whose paths curves near the surface. The vertical band structure in the 10\% sample correspond to the discrete trigger beam directions. Right: the full 10\% sample with the color axis referring to the maximum value of the maximum correlation for that bin. The events with high correlation are clustered in the top left hand corner.}
  \label{fig:DeepBox}
\end{figure*}
While typically a correlation map describes potential source locations in three dimensions, the Phased Array instrument records no azimuthal information due to the cylindrical symmetry around its z-axis. Instead, $R$-$z$ correlation maps are generated, from which many pairs of ($R$, $z$) can be identified as potential source locations. Due to the tight packing and small baselines of the phased array antennas, these maps generally do not have significant discerning power for the exact location of either $R$ or $z$ except for extremely bright and/or nearby signals such as the local calibration pulser. One exception to this is events which contain both direct and refracted pulses; an example of the correlation map for a simulated event with two pulses is shown in Figure \ref{fig:correlationMap}. The additional A5 baseline channels can be used to reconstruct the azimuthal direction.

For each map, at the location where the correlation value is largest, the corresponding time delays can be used to add the signals together with appropriate time delays to generate a single waveform, called the coherently summed waveform (CSW). From the CSW, analysis variables such as SNR, peak power, and impulsivity are computed which are used in the analysis to distinguish neutrino-like signals from noise. We describe calculations of these variables in a later section.

\section{Event Selection}
\subsection{Defining a Deep Region}
In addition to thermal noise fluctuations, which make up the vast majority of background, there are backgrounds that share characteristics with the expected neutrino signal. The most challenging backgrounds to remove are anthropogenic backgrounds and cosmic ray interactions, which can both look impulsive, isolated in time, and have strong peaks in their correlation maps.

The biggest difference between these backgrounds and neutrino candidates is the incoming direction: anthropogenic backgrounds and cosmic ray interactions both occur at or near the surface, while the neutrinos we expect to see mostly interact in deep ice. We define a phase space in which cosmic and anthropogenic backgrounds associated with surface locations are well separated from the neutrino-induced events, referred to as the signal region or the deep region. Two variables were used to define this deep region: zenith angle reconstructed at the antenna array, and the Surface Correlation Ratio, the ratio of the maximum correlation value within 10~m of the surface to the global maximum correlation value, each retrieved from the values in the $R$-$z$ correlation map. By plotting the 10\% sample using these coordinates, the highly correlating events that are likely a combination of anthropogenic and cosmic ray backgrounds cluster at high Surface Corrleation Ratio and low Zenith angle, as shown in Figure \ref{fig:DeepBox}. 

The boundary of the deep region in reconstructed zenith angle is chosen to be 57~degrees, calculated using Snell's law to find the arrival zenith angle for a signal that experienced a total internal reflection at 20~m below the surface. This boundary is chosen to exclude cosmic rays, which are expected to emit radio waves at a depth of less than 10 m \cite{deVries_CR_2016}. The boundary in the Surface Correlation Ratio is chosen to be 0.5, so that only events correlating twice as well below 10~m were kept in the signal region. All other events were sorted into the surface region, which is not part of the signal region. The requirement for events to be located in the signal region removed 21\% of simulated neutrinos, most of which were neutrinos that interacted deep in the ice, yet their signals traveled on a refracted path up near the surface before curving back down to the detector. It is possible that a future analysis could recover some of these simulated events by fully utilizing the baseline ARA antennas.

\subsection{Non-Thermal Background Removal and Data Cleaning}
While the vast majority of events recorded by the Phased Array are expected to be thermal fluctuations produced by the ice, there are additional backgrounds that must be removed from the sample before characterizing the thermal noise distribution. Those backgrounds are:

\textbf{Calibration Pulser Events}: During a typical Phased Array data run, the local calibration pulser transmits a VPol pulse at a rate of 1~Hz on the second as determined by GPS. These events are removed from the sample based on their trigger time, which only removes 0.016\% of simulated neutrinos. While all calibration pulses in the 10\% sample are successfully removed based on their trigger time alone, analyses of other stations showed that 1~out of 10,000~calibration pulses did not occur within the expected time window. As a precaution, events reconstructing within a box defined as between 45-55 m radially from the antenna array and between 171-175.5~m deep are also removed from the analysis. The 90\% upper limit of expected calibration pulser events occurring outside of the defined box is calculated to be 21 events; with the geometry-based metric, 99.8\% of calibration pulser events were expected to be removed, resulting in a 90\% upper limit on the background estimate of 0.009 and a simulation efficiency of 99.64\%. After unblinding, we find that all calibration pulser events are removed by the timing requirement, suggesting that future analyses for this station may not need the added geometry requirement.

\textbf{Software Triggered Events}: The Phased Array generates internal software triggers by randomly sampling the noise environment at a rate of 1~Hz, to check that the system is operating normally. These events are used to measure the average root-mean-square of the noise for each channel over all events in each run, which is used in calculations of analysis variables such as the SNR. The software triggered events are tagged by the data acquisition software and removed from the analysis.

\textbf{Continuous Wave (CW) Backgrounds}: CW contamination in our data is caused by anthropogenic sources visible to A5 and is identifiable in triggered events as a peak in the measured spectra. The most prevalent CW source is from radio communications among station personnel, which is transmitted at a frequency of 450~MHz, near the center of the Phased Array's frequency band. To remove this, our instrument is equipped with a notch filter at the hardware level. 
\begin{table*}
\begin{center}
 \begin{tabular}{||c c c||} 
 \hline
 Variable Name & Variable Description & Variable Range   \\ [0.5ex] 
 \hline\hline
Maximum Correlation & Maximum value on the $R$-$z$ correlation map & [-1.0, 1.0] \\ 
 \hline
Best R & Location of the maximum correlation (lateral distance, m) & [10~m, 5500~m] \\
 \hline
Best Z & Location of the maximum correlation (depth, m) & [-1500~m, 0~m] \\
 \hline
Hilbert Peak & Magnitude of the peak of the coherently summed waveform (CSW) & [0, 63]  \\
 \hline
Surface Correlation Ratio & Maximum correlation within 10~m of the surface & [-1.0, 1.0]\\ 
 \hline
SurfaceZ & Location of SurfaceCor (depth, m) & [-10~m, 0~m]\\ 
 \hline
SurfaceR & Location of SurfaceCor (lateral distance, m) & [10~m, 5500~m]\\ 
 \hline
CloseSurface & Maximum value of correlation map above the surface & [-1.0, 1.0]\\ 
 \hline
Zenith Angle & Best reconstructed zenith angle, calculated at the antenna array & [0, 180.0]\\ 
 \hline
WindowedZenith & Best reconstructed zenith angle, windowed around the trigger & [0, 180.0]\\ 
 \hline
CoherentSNR & SNR of the CSW & Not Constrained\\ 
 \hline
AvgSNR & Average of the SNR of each individual VPol waveform & Not Constrained \\ 
 \hline
Impulsivity & Average of the cumulative distribution function (CDF) around the peak of the CSW & [0, 1.0]\\ 
 \hline
$R^2$ & Correlation Coefficient for the linear fit to Impulsivity CDF & [0, 1.0]\\ 
 \hline
Slope & Slope of linear fit to impulsivity CDF & Not Constrained\\ 
 \hline
Intercept & Intercept of linear fit to impulsivity CDF & Not Constrained\\ 
 \hline
PowerSpot & Location of peak power along the CSW &  [0,2560]\\ 
 \hline
A5 Correlation & Maximum correlation between Phased Array CSW and A5 channel & [-1.0, 1.0]\\ 
 \hline
 K-S & \begin{tabular}{@{}c@{}}The Kolmogorov–Smirnov (K-S) test, \\ comparing the impulsivity CDF to a linear hypothesis\end{tabular}   & Not Constrained \\[1ex] 
 \hline
 
\end{tabular}
\caption{Table of analysis variables used in the Fisher Discriminant, along with their definitions and expected ranges where applicable. For more complete descriptions, please see the Appendix.}
\label{table:1}
\end{center}
\vspace{-10pt}

\end{table*}
There are two additional sources of CW that are more sporadic and must be removed in analysis. The first source is weather balloons, which are launched twice per day and communicate back to South Pole Station at a frequency of 405~MHz. The second source is satellites with downlink frequencies around 137~MHz; although this is technically outside of the Phased Array's band, it does cause the trigger rates, and therefore thresholds, of the trigger beams to cycle up and down as satellites come into and out of view. Both of these frequencies are targeted with a sine subtraction filter developed by ANITA \cite{ANITA_III}, where a frequency was filtered if removing the frequency decreased the power by at least 4\%, a threshold chosen for this analysis based on CW-contaminated events within the 24-hour unblinded sample. While no other CW sources were visible in the 10\% sample, an additional sine subtraction filter was applied to all events over the entire frequency range with a power threshold of 10\%, in case other transient CW sources, such as satellites at other frequencies, appeared in the full data sample.

\begin{figure*}[htp]
  \centering
  {\includegraphics[width=0.48\textwidth]{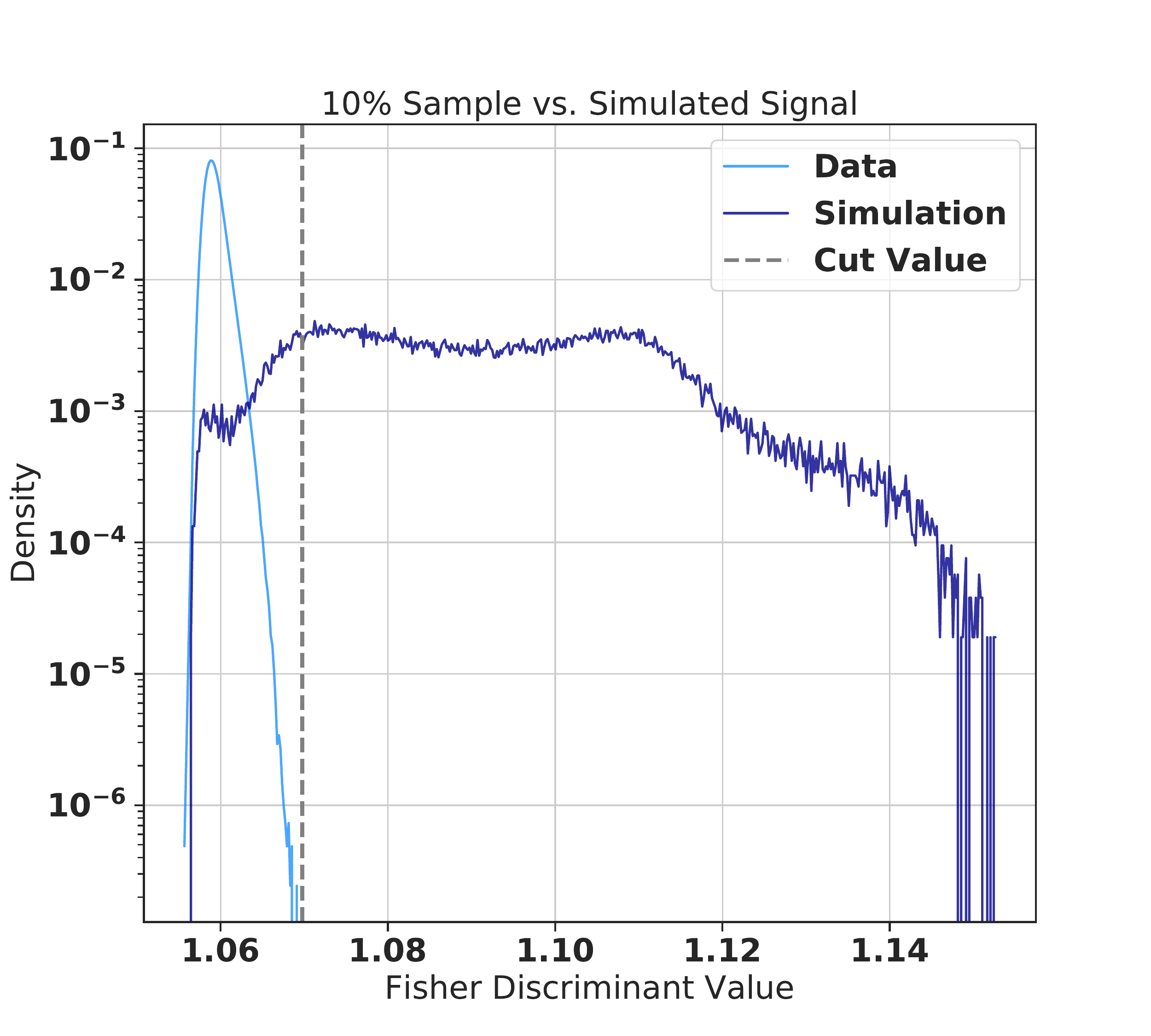}\label{fig:f1}}
  \includegraphics[width=0.48\textwidth]{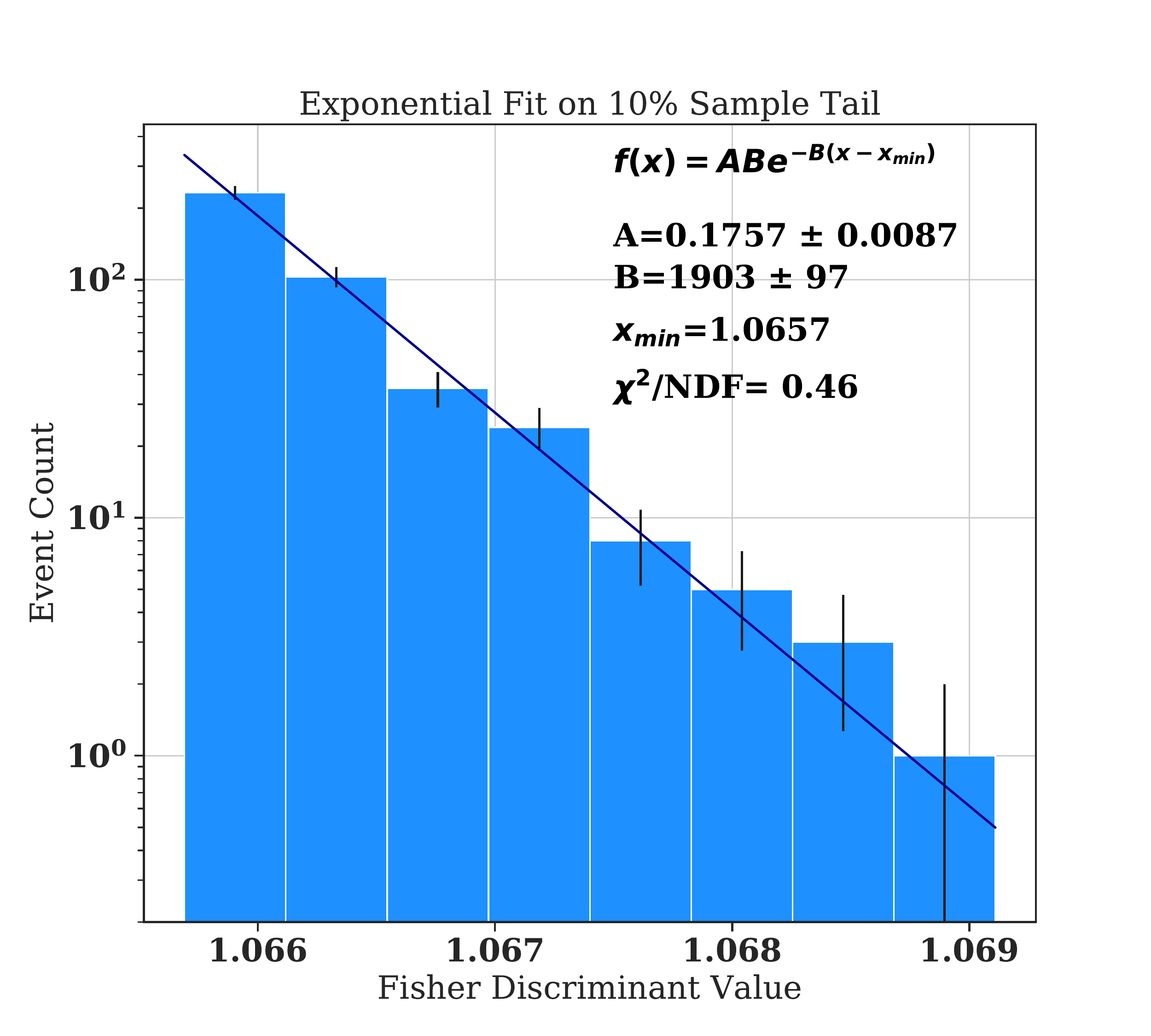}\label{fig:f2}
  \caption{Left: A comparison between the 10\% sample distribution and the simulated neutrino flux, with the optimized threshold marked with a vertical line. Right: the tail of the 10\% sample, with the exponential fit overlaid. The value of $X_{min}$ is defined as the minimum value of the Fisher Discriminant in the tail of the 10\% sample.}
\label{fig:fisher}
\end{figure*}

\textbf{Additional Livetime Cuts}: Several time periods were eliminated due to station configuration changes or known station activity, totalling 57~hours. Of this, 18~hours were removed due to multiple surface events occurring within a single run, common during known station activity, including remote configuration changes. Additionally, 39~hours of data were removed due to the calibration pulser rate dropping; this is most common when the calibration pulser is switched from one mode to another, e.g. VPol pulsing to HPol, or GPS-tied pulses to continuous noise. During these mode switches, spurious signals are possible, which would be challenging to remove in analysis. In total, this removed 2.3 days of livetime, leaving 206.4~days of data that can be used in analysis.

\subsection{Fisher Discriminant}

After applying the quality cuts, the distributions of the analysis variables of the remaining 10\% unblinded sample shares the characteristics expected of thermal-noise-triggered events. To distinguish between this thermal-noise-like sample and expected neutrino signals, a Fisher discriminant \cite{Fisher} was trained on the remaining 10\% sample and simulated neutrinos. The input variables for the Fisher discriminant are listed in Table \ref{table:1}, and the resulting distributions of data and simulation are seen in Figure~\ref{fig:fisher}.

The Fisher Discriminant cut was chosen by optimizing for the best sensitivity, defined as the mean 90\% upper limit using the Feldman-Cousins method \cite{Feldman:1997qc} divided by the analysis efficiency. This optimization was carried out including uncertainties using the following method. First, the tail of the 10\% sample was fit to an exponential, with uncertainties on each fit parameter calculated using the covariance matrix. The resulting fit can be seen in Figure~\ref{fig:fisher}. Next, for each potential cut value, 100,000~Monte-Carlo psuedo-experiments were run, with fit parameters varied by sampling from Gaussian distributions defined by the fit parameters and their uncertainties. Then, for each set of sampled parameters, the background was calculated by integrating the fit function from the cut value outward and scaling by nine to account for the increased livetime of the full 90\% sample. The median of this background distribution was added to the total background from other cuts and used to calculate the mean upper limit from 500~psuedo-experiments. In this way, the expected sensitivity, as well as the expected background distribution, can be determined for all possible cuts on the Fisher discriminant. The background that optimizes the expected sensitivity for the 90\% sample is $0.09_{-0.04}^{+0.06}$; because the distribution of the mean number of background events is asymmetric, the median, 16th percentile, and 84th percentile are reported. In the 10\% sample, 0~events passed all cuts in the deep region. 
\begin{table*}
\begin{center}
 \begin{tabular}{||c c c c||} 
 \hline
 Cut Name & ~~~Events Remaining~~~ & ~~~Background Estimate~~~ & Signal Efficiency   \\ [0.5ex] 
 \hline\hline
 None & 18,651,857 & N/A~& 100\% \\ 
 \hline
 Deep Region Boundary & 6,005,122 & N/A & 79.03\% \\
 \hline
 Cal Pulser Gate Flag & 4,423,436 & N/A & 99.98\% \\
 \hline
 Cal Pulser Geometry Cut & 4,411,686 & 0.009 & 99.64\% \\
 \hline
 Software Trigger Cut & 4,014,776 & N/A & 100\% \\
 \hline

 Fisher Discriminant & 0 & $0.09_{-0.04}^{+0.06}$ & 86.58\% \\ 
 \hline

 Total & 0 & $0.10_{-0.04}^{+0.06}$ & 68.16\% \\[1ex] 
 \hline
\end{tabular}
\caption{Table of cuts, background estimates, and analysis efficiencies from the 10\% sample in the deep region. For the Fisher Discriminant, the median background is reported, as are the 16th and 84th percentiles.}
\label{table:cuts}
\end{center}
\vspace{-10pt}

\end{table*}
\section{Results} 

\begin{figure*}[!btp]
  \centering
  \includegraphics[width=0.48\textwidth]{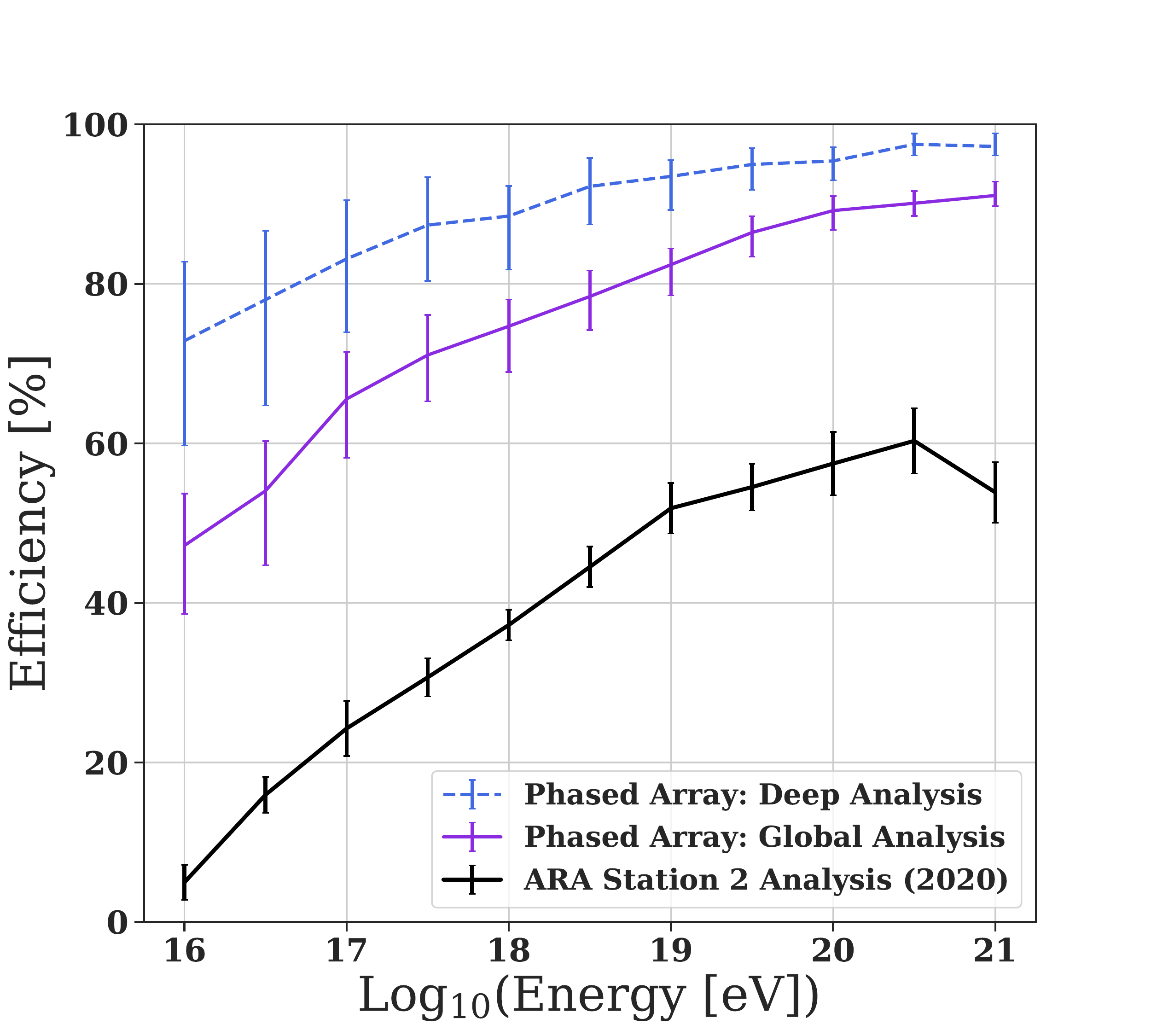}\label{fig:Energy_eff}
  \includegraphics[width=0.48\textwidth]{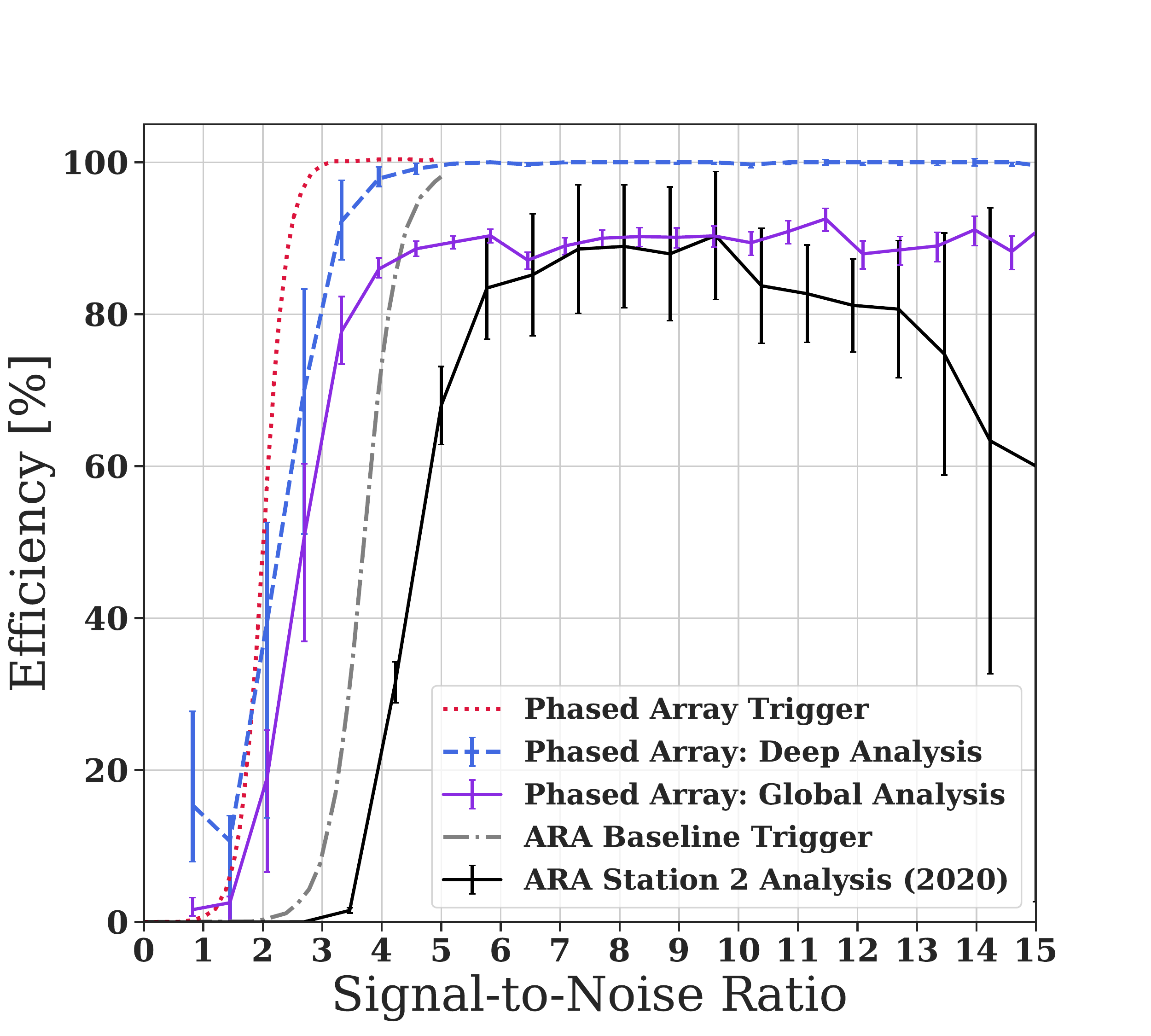}\label{fig:SNR_eff}
  \caption{Analysis efficiency of this analysis compared to the most recent ARA Station 2 analysis efficiency \cite{ARA23}. The blue curve is the efficiency on neutrinos in the deep region; the purple curve includes the efficiency loss from removing the surface region which results in a 21\% loss in efficiency averaged over all energies. The analysis efficiency shown in the plot on the left is generated for each energy bin separately. The analysis efficiency shown in the plot on the right is generated based on a cosmogenic flux \cite{Kotera:2010yn}.The trigger efficiency vs. SNR is from \cite{PhasedArrayInstrument}, while the analysis efficiency from ARA Station 2 is adapted from \cite{ARA23}.}
 \label{fig:efficiencies}
\end{figure*}

\subsection{Deep Region Results}
\begin{figure*}[ht]
\centering
\includegraphics[width=\textwidth]{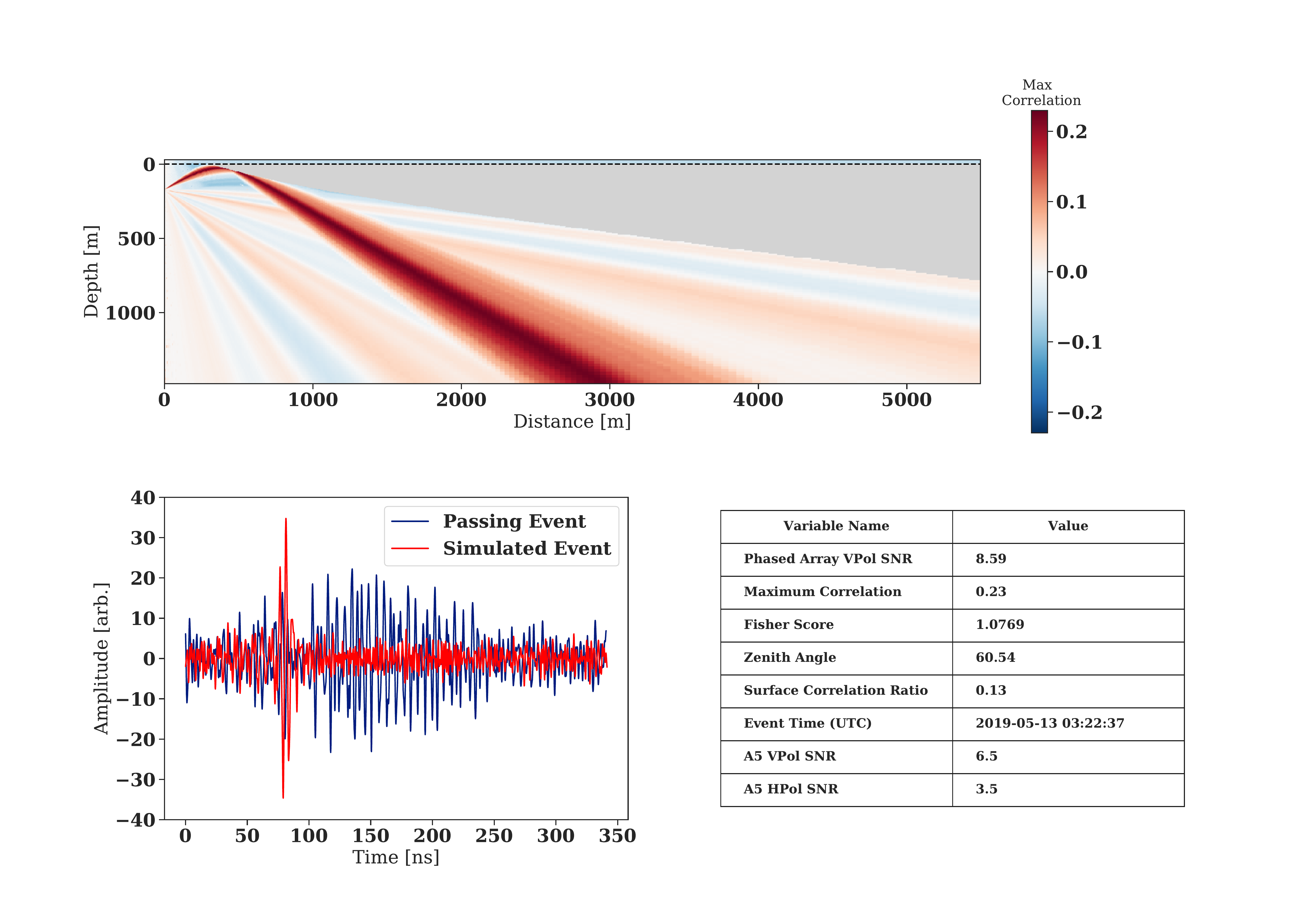}
\caption{Top: the correlation map of the passing event. Bottom left: the Coherently Summed Waveform of the passing event, compared to a simulated event at the same incoming angle. The noise in the simulated waveform is normalized to have the same root-mean-square as the average from data. Bottom right: a table showing some of the values of analysis variables for the passing event.}
\label{fig:passingEvent}
\end{figure*}

\begin{figure}[ht]
\centering
\includegraphics[width=\columnwidth]{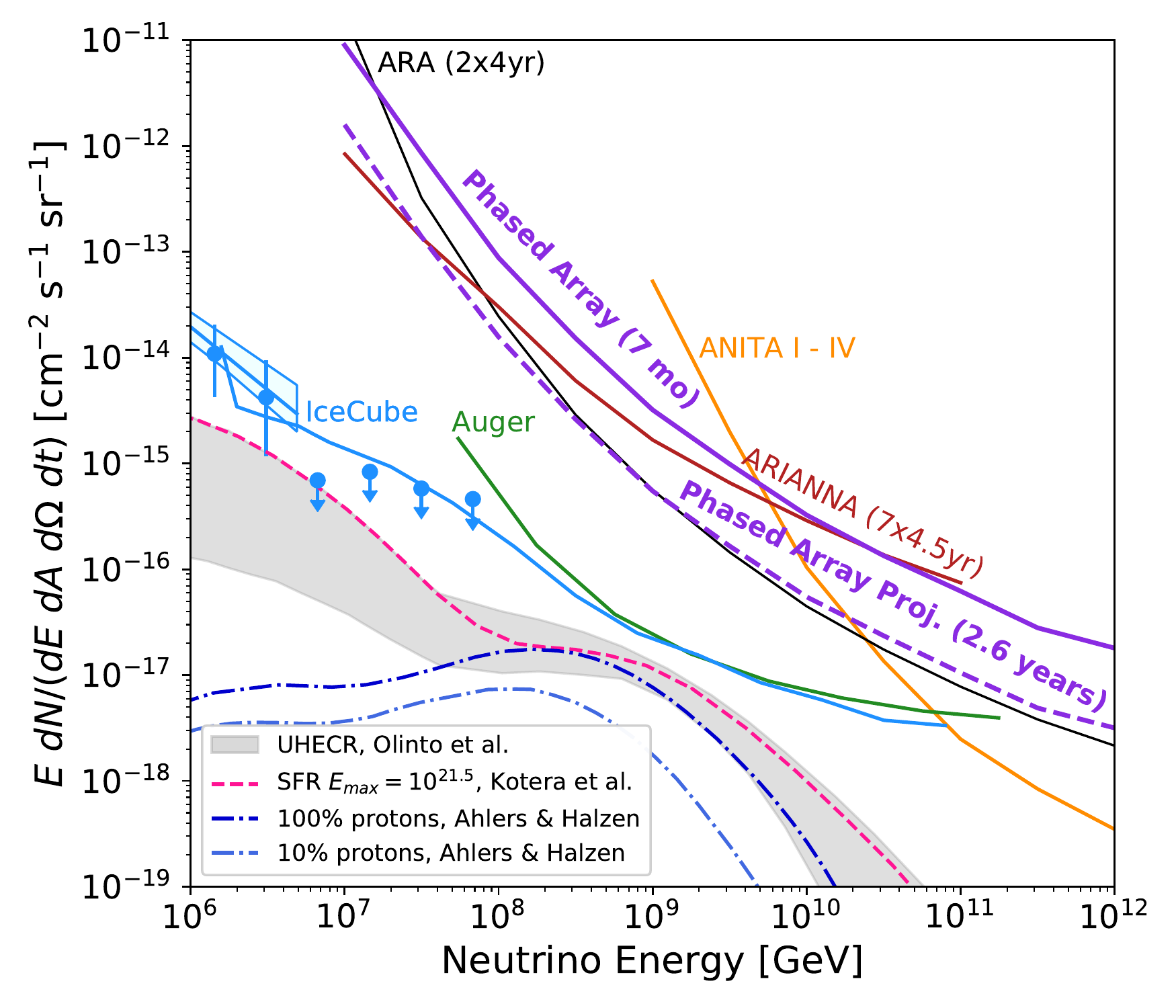}
\caption{The limit from this Phased Array analysis, using one station and six months of livetime, along with the projected sensitivity from the entire available livetime of 2.6~years. The expected sensitivity is calculated by directly scaling the livetime and expected backgrounds, selecting the median expected background, and calculating the 90\% upper limit. Plotted for comparison are the previous published results from ARA (two stations, each with 4.0 station-years), along with results from ANITA \cite{ANITA_IV}, ARIANNA (seven stations, each with 4.5 station-years) \cite{ARIANNA}, Auger \cite{Auger} and IceCube\cite{IceCube_limit1} \cite{IceCube_limit2}. Additionally, theoretical production models of cosmogenic neutrinos are plotted \cite{Olinto:2011ng} \cite{Kotera:2011cp} \cite{Ahlers:2012rz}.  }
\label{fig:limitplot}
\end{figure}

The final cuts and background estimate for the deep region are shown in Table \ref{table:cuts}. Additionally, in Figure \ref{fig:efficiencies} we report the analysis efficiency as a function of neutrino energy and as a function of SNR. For comparison, our efficiency is compared against the most recent published ARA result, ARA Station 2, as the baseline ARA5 data is not within the scope of this paper and will instead be analyzed in a future work. The sideband region just outside the deep signal region (52-57~degrees in zenith) was unblinded first as a validation of our background estimate and was not part of the signal region. In this region, we observe zero events passing all cuts. After unblinding the deep region only, we observe one event on a background of $0.10_{-0.04}^{+0.06}$. This event was triggered separately by both the baseline ARA trigger and the Phased Array trigger; its p-value is calculated to be 0.11 using Poisson statistics. The event is included in the limit set in Figure~\ref{fig:limitplot} and is shown in detail in Figure~\ref{fig:passingEvent}.

Various hypotheses for the origin of the single passing event have been investigated. As discussed in \cite{triboelectric_2021}, high wind speeds can cause a noticeable radio background; the wind speed at the time of this event was 2 m/s, well below the average of 6~m/s at the station location. The event occurred during the austral winter and reconstructs in a direction pointing away from South Pole Station, making anthropogenic backgrounds unlikely but not impossible. Because the baseline A5 trigger also saw this event, we can rule out a glitch event internal to the Phased Array DAQ. 

Neutrinos simulated using the ice model from Equation \ref{eq:1} with similar arrival angles are almost exclusively refracted neutrino events, in which the radio signal originates deep in the ice but curves near the surface before hitting the detector. The passing event is a longer, brighter signal than most simulated neutrino events, as seen in Figure \ref{fig:passingEvent}. 

\subsection{Surface Region Analysis}
The surface region was not included in the neutrino search in the analysis, due to an expected background of cosmic ray showers and anthropogenic noise. After completing the neutrino analysis, the surface region was unblinded and the same cuts were applied to this region. A total of 46~events passed the cuts. Of those events, 7 were not impulsive and reconstructed poorly; additionally, 3 events were time-clustered and likely anthropogenic. The remaining 36~were uniformly distributed in time, reconstructed to the surface well, and were generally impulsive, suggesting the sample likely contains cosmic ray candidates. By using the full A5 instrument to calculate azimuth angle, there was no excess of events that pointed towards South Pole Station. As discussed in \cite{CosmicRay_Askaryan}, we expect multiple types of cosmic ray signals, including both in-air showers and impacting ice core events. More study and discussion of potential cosmic ray candidates will be done in future publications.

\section{Discussion} 
\subsection{Limit on the diffuse neutrino flux and projected expected sensitivity}

Using the signal region as defined above, with 1 event found in the 90\% sample on a background of $0.10_{-0.04}^{+0.06}$, the 90\% upper limit was calculated using the method laid out in \cite{ARA23} and is shown in Figure \ref{fig:limitplot}. The key pieces of this calculation are: (1) the Feldman-Cousins upper limit; (2) the livetime; (3) the effective area, calculated from the effective volume and the interaction length of a neutrino interacting in the Earth; and (4) the analysis efficiency. This limit is the most sensitive published limit from the ARA collaboration at the lowest end of ARA's targeted energy range, even as the livetime of this analysis is lower by a factor of 14 compared to previous analyses. This improvement is a function of both a more sensitive trigger, leading to a higher effective volume, and improved analysis techniques made possible by the simplicity of the phased array geometry, leading to better zenith angle reconstruction and a larger signal region. 

While the lower trigger threshold afforded by the phased-array technique is well-established, there has been an open question about whether the lower-amplitude events afforded by this technique would be efficiently separated from background. In this work we show that they can be and that the lower trigger threshold does indeed result in improved neutrino sensitivity at the analysis level. This is an optimistic finding for the other experiments planning to capitalize on a Phased Array trigger design to improve effective volumes \cite{RNOG_2021} \cite{PUEO_white} \cite{Wissel_2020} \cite{hallmann2021sensitivity}.

As we scale up to larger experiments with longer livetimes and larger effective volumes, it is imperative to reduce backgrounds even further, and and it will be important to understand the origin of the single passing event from this analysis, which is unlikely to be thermal noise. Additionally, if the single passing event were a neutrino signal, the flux would be approximately 100 times greater than the IceCube limit at energies above 10~PeV, considering the small area of ice surveyed, the small livetime included, and the higher energy threshold of the Phased Array compared with IceCube. This suggests a new type of background that was not modeled in this analysis and should be studied further. One potential explanation is a cosmic ray impacting core event penetrating the ice deeper than expected, or traveling through a poorly modeled region of ice near the surface. This hypothesis is helped by the fact that the trajectory intersects the ``shadow" region close to the surface \cite{ShadowRegion_Deaconu}. Better modeling of the ice near the shadow region, as well as ARA-specific cosmic ray simulations, would both help us understand the origin of this event, and help recover a larger fraction of the surface region to be used in a neutrino search.

Although this analysis focused on data taken in 2019, the Phased Array currently has just under 1000~days of total livetime available, with more data taken every year. However, future analyses that take advantage of this data set will need to make some assumptions about the passing event. The projection in Figure \ref{fig:limitplot} takes the pessimistic assumption that only events reconstructing below 90 degrees are usable; the efficiency as a function of zenith angle is included in the Appendix in Figure \ref{fig:ZenithEfficiency} for reference. In practice, an analysis on this full data could take advantage of a combination of characteristics that make this event an anomaly: its high SNR with relatively low correlation, its unusually long waveform, or its zenith angle near the edge of the signal region. Even considering the pessimistic option, the Phased Array alone could likely achieve the greatest sensitivity for the ARA experiment thus far across much of the energy range. Additional benefits can come from analyzing the data from the baseline ARA instrument, as well as the data taken from the other four ARA stations. 

The high analysis efficiency in the deep region in particular is also encouraging. While the surface region was not included as part of the signal region in this analysis, the methods used in the deep region can be applied to the surface as well, provided that the types of expected backgrounds in the surface region are understood. Alternatively, the size of the deep region could be increased by taking advantage of additional information provided by the baseline ARA instrument.

\subsection{Systematic Uncertainties} \label{uncertainties}

The limit calculation includes uncertainties using the method described in \cite{Conrad:2002kn}. In particular, we consider the uncertainties on the effective volume calculation and the analysis efficiency.

The uncertainties on the effective volume are introduced by the AraSim simulation package and are plotted in Figure \ref{fig:EffectiveVolume}. Uncertainties in the neutrino cross section, the Askaryan emission model, and the attenuation length of ice are explicitly laid out in \cite{ARA23}, which uses the same simulation code as we do in this work, allowing the uncertainties to directly carry over. 

The key differences between the simulation used in \cite{ARA23} and this work are the ice model, the signal chain, and the trigger efficiency. The ice model is parameterized in the form $n(z)=A-B e^{C z}$, with the parameters found to have the following uncertainties: $A=1.780 \pm 0.005$, $B=0.454 \pm 0.084$, $C= -0.0202 \pm 0.0004 \mathrm{m}^{-1}$. While the model is different than the model used in previous ARA analyses \cite{ARA23}, two of the three parameters are within one standard deviation across the models, with the third being better constrained in the model used here due to the short baselines of the Phased Array. We estimate that the uncertainty in our ice model introduces a 5\% uncertainty in the effective volume. Even though the ice model was fit using calibration pulses from deep ice, all surface calibration events were successfully pointed back at the sources using the new ice model. 

The signal chain uncertainty is dominated by the uncertainties in the antenna response, described in \cite{Allison:2015eky} and estimated as a 10\% effect in total. The Phased Array signal chain uncertainty independent of the antennas is measured in the lab as $< 5\%$. From these measurements, we estimate that with the antennas included, the total signal chain uncertainty for the Phased Array instrument is also 10\%. From \cite{ARA23}, this adds a 3\% uncertainty in the effective volume.

The Phased Array trigger is simulated separately from the typical ARA trigger, and has its own uncertainties. Unlike the previous analysis, the output of the Phased Array simulation was tuned to perfectly match the trigger efficiency as a function of SNR. In \cite{PhasedArrayInstrument}, we estimate that the error on the location of the 50\% trigger point used in the simulation is $1.8 \pm 0.2$ in SNR. This translates to a total trigger uncertainty of approximately $\pm 5$ \% for events with SNR below 3; this corresponds to a $\pm 2.5 \%$ impact on effective volume at $10^{16}$ eV and decreases to $\pm 1 $\% or less at energies above $10^{18.5}$ eV.

Finally, we consider the uncertainties related to the analysis, introduced by the Fisher Discriminant. By calculating the efficiency on simulated neutrinos for the expected background range, the uncertainty on the simulation efficiency is found to be $68\% + 5\% - 3\%$ globally. Events with the lowest calculated SNRs are most impacted by this, as they are more likely to be assigned Fisher Discriminant values similar to noise.

\section{Conclusion}
The results presented here represent the most efficient analysis conducted with the ARA instrument, and the lowest threshold analysis ever conducted out of all radio-detection neutrino experiments. We show that the gains made possible by a more efficient trigger are gains that can be carried through the analysis, motivating the design of future Phased Array triggers for in-ice neutrino detectors. Finally, further modeling of the backgrounds in the surface region, as well as investigations into the origins of the passing event, may lead to additional gains in analysis efficiency.

\section*{Acknowledgments}

Kaeli Hughes was the main author of this manuscript and led the analysis discussed. The ARA Collaboration designed, constructed, and now operates the ARA detectors. We would like to thank IceCube and specifically the winterovers for the support in operating the detector.  Data processing and calibration, Monte Carlo simulations of the detector and of theoretical models and data analyses were performed by a large number of collaboration members, who also discussed and approved the scientific results presented here.  We are thankful to the Raytheon Polar Services Corporation, Lockheed Martin, and the Antarctic Support Contractor for field support and enabling our work on the harshest continent.  We are thankful to the National Science Foundation (NSF) Office of Polar Programs and Physics Division for funding support.  We further thank the Taiwan National Science Councils Vanguard Program NSC 92-2628-M-002-09 and the Belgian F.R.S.-FNRS Grant 4.4508.01. K.~Hughes thanks the NSF for support through the Graduate Research Fellowship Program Award DGE-1746045. B.~A.~Clark thanks the NSF for support through the Astronomy and Astrophysics Postdoctoral Fellowship under Award 1903885, as well as the Institute for Cyber-Enabled Research at Michigan State University. A.~Connolly thanks the NSF for Award 1806923 and also acknowledges the Ohio Supercomputer Center. S.~A.~Wissel thanks the NSF for support through CAREER Award 2033500.  D.~Besson acknowledges support from National Research Nuclear University MEPhi (Moscow Engineering Physics Institute). A.~Vieregg thanks the Sloan Foundation and the Research Corporation for Science Advancement, the Research Computing Center and the Kavli Institute for Cosmological Physics at the University of Chicago for the resources they provided.  R. Nichol thanks the Leverhulme Trust for their support. K.D.~de~Vries is supported by European Research Council under the European Unions Horizon research and innovation program (grant agreement 763 No 805486). D.~Besson, I.~Kravchenko, and D.~Seckel thank the NSF for support through the IceCube EPSCoR Initiative (Award ID 2019597). 

\bibliographystyle{bst/apsrev4-1}
\bibliography{references}

\clearpage
\newpage

\section{Appendix}

\begin{figure*}[b!]
\centering
\includegraphics[width=0.95\textwidth]{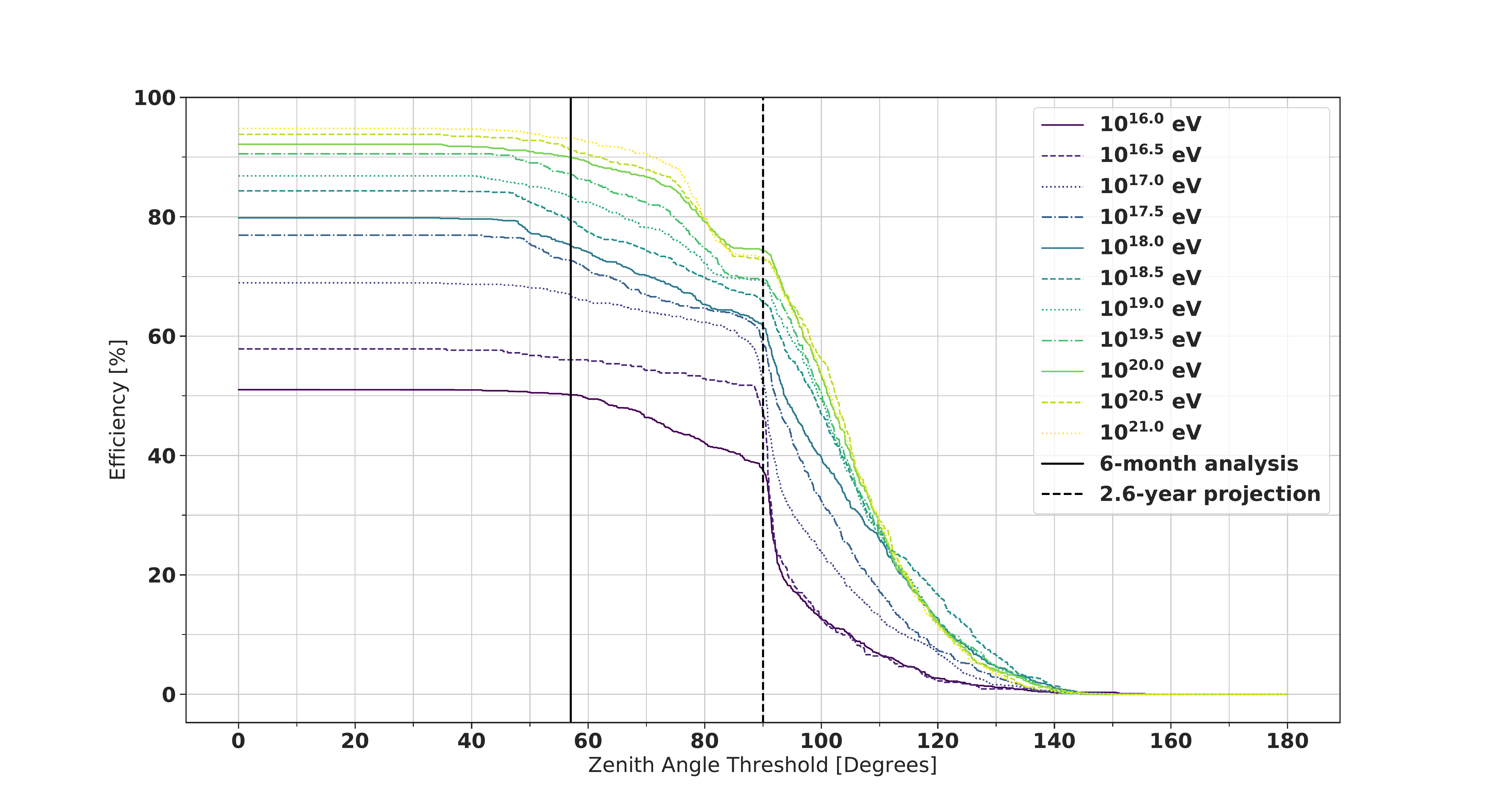}
\caption{The efficiency of the analysis as a function of zenith angle and half-decade energy bin. The solid black line shows where the zenith cut was applied for this analysis; the dashed black line is used for the projection in Figure \ref{fig:limitplot}.}
\label{fig:ZenithEfficiency}
\end{figure*}

\subsection{Calculation of Impulsivity Measures}
For each event in the analysis, three correlation maps are generated, and a CSW is created, as discussed in Section \ref{IceModel}. From this, the impulsivity $I$ is calculated using the same method as \cite{ANITA_III}, where the CDF of the power of the waveform is generated from the location of the peak power outwards. For a noise event, this CDF should be nearly linear, as the power is distributed evenly throughout the CSW; for an impulsive event, the CDF should be steeply rising at the beginning and nearly flat towards the end. The impulsivity $I$ is defined as $2A-1$, where $A$ is the average of the CDF.

Additionally, because the CDF is generally linear in the case of a noise event, other metrics can also be calculated from the CDF. A least-squares fitting method is used to calculate a linear fit for the CDF, and the slope, intercept, and the correlation coefficient are used as analysis variables. Additionally, a K-S test statistic is calculated as the largest distance between the linear fit and the CDF.

\subsection{Effect of Zenith Cut on Efficiency}
Because this analysis includes a passing event at a zenith angle of 60.5\degree, future analyses may want to consider how such an event could be removed. Possibly the simplest solution is a requiring all events to reconstruct at an angle greater than 60.5\degree, although likely a more targeted approach would be more efficient.

The projected limit in Figure \ref{fig:limitplot} takes a pessimistic approach and includes the efficiency hit if a zenith threshold at 90\degree is applied in the analysis. Other options could include a zenith threshold at $79\degree$, corresponding to the most pessimistic zenith angle used in previous ARA analyses \cite{ARA23}, or a zenith threshold at $61\degree$, just greater than the zenith angle of the passing event. For the convenience of the reader, the efficiency of all zenith angle thresholds as a function of energy is included in Figure \ref{fig:ZenithEfficiency}.

\end{document}

%% file: ara_revtex_institutes.tex

\newcommand{\atOSU}{\affiliation{Dept. of Physics, Center for Cosmology and AstroParticle Physics, The Ohio State University, Columbus, OH 43210}}
\newcommand{\atChiba}{\affiliation{Dept. of Physics, Chiba University, Chiba, Japan}}
\newcommand{\atUW}{\affiliation{Dept. of Physics, University of Wisconsin-Madison, Madison,  WI 53706}}
\newcommand{\atKU}{\affiliation{Dept. of Physics and Astronomy, University of Kansas, Lawrence, KS 66045}}
\newcommand{\atMoscow}{\affiliation{Moscow Engineering Physics Institute, Moscow, Russia}}
\newcommand{\atNTU}{\affiliation{Dept. of Physics, Grad. Inst. of Astrophys., Leung Center for Cosmology and Particle Astrophysics, National Taiwan University, Taipei, Taiwan}}
\newcommand{\atMSU}{\affiliation{Dept. of Physics and Astronomy, Michigan State University, East Lansing, Michigan 48824}}
\newcommand{\atUC}{\affiliation{Dept. of Physics, Enrico Fermi Institue, Kavli Institute for Cosmological Physics, University of Chicago, Chicago, IL 60637}}
\newcommand{\atUCL}{\affiliation{Dept. of Physics and Astronomy, University College London, London, United Kingdom}}
\newcommand{\atULB}{\affiliation{Universite Libre de Bruxelles, Science Faculty CP230, B-1050 Brussels, Belgium}}
\newcommand{\atVUB}{\affiliation{Vrije Universiteit Brussel, Brussels, Belgium}}
\newcommand{\atUMD}{\affiliation{Dept. of Physics, University of Maryland, College Park, MD 20742}}
\newcommand{\atWhittier}{\affiliation{Dept. Physics and Astronomy, Whittier College, Whittier, CA 90602}}
\newcommand{\atUD}{\affiliation{Dept. of Physics, University of Delaware, Newark, DE 19716}}
\newcommand{\atPSUigc}{\affiliation{Center for Multi-Messenger Astrophysics, Institute for Gravitation and the Cosmos, Pennsylvania State University, University Park, PA 16802}}
\newcommand{\atPSUphys}{\affiliation{Dept. of Physics, Pennsylvania State University, University Park, PA 16802}}
\newcommand{\atUNL}{\affiliation{Dept. of Physics and Astronomy, University of Nebraska, Lincoln, Nebraska 68588}}
\newcommand{\atWeizman}{\affiliation{Weizmann Institute of Science, Rehovot, Israel}}
\newcommand{\atDenison}{\affiliation{Dept. of Physics and Astronomy, Denison University, Granville, Ohio 43023}}
\newcommand{\atUH}{\affiliation{Dept. of Physics and Astronomy, University of Hawaii, Manoa, HI 96822}}
\newcommand{\atPSUast}{\affiliation{Dept. of Astronomy and Astrophysics, Pennsylvania State University, University Park, PA 16802}}
\newcommand{\atCalPoly}{\affiliation{Physics Dept., California Polytechnic State University, San Luis Obispo, CA 93407}}
\newcommand{\atSHINE}{\affiliation{SHINE Technologies, 3400 Innovation Dr., Janesville, Wisconsin 53546, US}}

%% file: ara_revtex_authors.tex

 \author{P.~Allison}\atOSU
 \author{S.~Archambault}\atChiba
 \author{J.J.~Beatty}\atOSU
 \author{D.Z.~Besson}\atKU\atMoscow
 \author{A.~Bishop}\atUW
 \author{C.C.~Chen}\atNTU
 \author{C.H.~Chen}\atNTU
 \author{P.~Chen}\atNTU
 \author{Y.C.~Chen}\atNTU
 \author{B.A.~Clark}\atMSU
 \author{W.~Clay}\atUC
 \author{A.~Connolly}\atOSU
 \author{L.~Cremonesi}\atUCL
 \author{P.~Dasgupta}\atULB
 \author{J.~Davies}\atUCL
 \author{S.~de~Kockere}\atVUB
 \author{K.D.~de~Vries}\atVUB
 \author{C.~Deaconu}\atUC
 \author{M.~A.~DuVernois}\atUW
 \author{J.~Flaherty}\atOSU
 \author{E.~Friedman}\atUMD
 \author{R.~Gaior}\atChiba
 \author{J.~Hanson}\atWhittier
 \author{N.~Harty}\atUD
 \author{B.~Hendricks}\atPSUigc\atPSUphys
 \author{K.D.~Hoffman}\atUMD
 \author{E.~Hong}\atOSU
 \author{S.Y.~Hsu}\atNTU
 \author{L.~Hu}\atNTU
 \author{J.J.~Huang}\atNTU
 \author{M.-H.~Huang}\atNTU
 \author{K.~Hughes}\email[K. Hughes: ]{kahughes@uchicago.edu}\atUC
 \author{A.~Ishihara}\atChiba
 \author{A.~Karle}\atUW
 \author{J.L.~Kelley}\atUW
 \author{K.-C.~Kim}\atUMD
 \author{M.-C.~Kim}\atChiba
 \author{I.~Kravchenko}\atUNL
 \author{R.~Krebs}\atPSUigc\atPSUphys
 \author{Y.~Ku}\atPSUigc\atPSUphys
 \author{C.Y.~Kuo}\atNTU
 \author{K.~Kurusu}\atChiba
 \author{H.~Landsman}\atWeizman
 \author{U.A.~Latif}\atKU\atVUB
 \author{C.-J.~Li}\atNTU
 \author{T.-C.~Liu}\atNTU
 \author{M.-Y.~Lu}\atUW
 \author{B.~Madison}\atKU
 \author{K.~Madison}\atKU
 \author{K.~Mase}\atChiba
 \author{T.~Meures}\atUW
 \author{J.~Nam}\atChiba
 \author{R.J.~Nichol}\atUCL
 \author{G.~Nir}\atWeizman
 \author{A.~Novikov}\atKU
 \author{A.~Nozdrina}\atKU
 \author{E.~Oberla}\atUC
 \author{J.~Osborn}\atUNL
 \author{Y.~Pan}\atUD
 \author{C.~Pfendner}\atDenison\atSHINE
 \author{N.~Punsuebsay}\atUD
 \author{J.~Roth}\atUD
 \author{D.~Seckel}\atUD
 \author{M.~F.~H.~Seikh}\atKU
 \author{Y.-S.~Shiao}\atNTU
 \author{A.~Shultz}\atKU
 \author{D.~Smith}\atUC
 \author{S.~Toscano}\atULB
 \author{J.~Torres}\atOSU
 \author{J.~Touart}\atUMD
 \author{N.~van~Eijndhoven}\atVUB
 \author{G.S.~Varner}\atUH
 \author{A.~Vieregg}\atUC
 \author{M.-Z.~Wang}\atNTU
 \author{S.-H.~Wang}\atNTU
 \author{Y.H.~Wang}\atNTU
 \author{S.A.~Wissel}\atPSUigc\atPSUphys\atPSUast\atCalPoly
 \author{C.~Xie}\atUCL
 \author{S.~Yoshida}\atChiba
 \author{R.~Young}\atKU
\collaboration{ARA Collaboration}\noaffiliation